\documentclass[fleqn,usenatbib]{mnras}

\usepackage{newtxtext,newtxmath}
\usepackage[T1]{fontenc}

\DeclareRobustCommand{\VAN}[3]{#2}
\let\VANthebibliography\thebibliography
\def\thebibliography{\DeclareRobustCommand{\VAN}[3]{##3}\VANthebibliography}

\usepackage{graphicx}	% Including figure files
\usepackage{amsmath}	% Advanced maths commands
\usepackage{gensymb}
\usepackage[separate-uncertainty=true]{siunitx}
\usepackage{threeparttable}
\usepackage[table]{xcolor}    % loads also »colortbl«
\usepackage{float}
\usepackage{stfloats}
\usepackage{caption}
\usepackage{subcaption}
\usepackage{soul}
\usepackage[normalem]{ulem} 

\DeclareSIUnit\parsec{pc}
\DeclareSIUnit\jansky{Jy}
\DeclareSIUnit\gauss{G}

\title[MeerKAT dwarf nova detections]{A MeerKAT survey of nearby dwarf novae: I. New detections}

\author[J. Kersten et al.]{
J.~Kersten$^{1}$\thanks{E-mail: j.kersten@astro.ru.nl},
E.~Körding$^{1}$,
P.~A.~Woudt$^{2}$,
P.~J.~Groot$^{1,3,4}$,
D.~R.~A.~Williams$^{5}$,
\newauthor
I.~Heywood$^{6,7,8}$,
D.~L.~Coppejans$^{9}$,
C.~Knigge$^{10}$,
J.~C.~A.~Miller-Jones$^{11}$,
G.~R.~Sivakoff$^{12}$,
R.~Fender$^{6,4}$
\\
$^{1}$Department of Astrophysics/IMAPP, Radboud University, P.O. Box 9010, 6500GL Nijmegen, The Netherlands\\
$^{2}$Inter-University Institute for Data Intensive Astronomy, Department of Astronomy, University of Cape Town, 7701 Rondebosch, Cape Town, South Africa\\
$^{3}$South African Astronomical Observatory, Observatory Road, Observatory, South Africa\\
$^{4}$Department of Astronomy, University of Cape Town, Private Bag X3, Rondebosch 7701, South Africa\\
$^{5}$Jodrell Bank Centre for Astrophysics, School of Physics and Astronomy, The University of Manchester, Manchester, M13 9PL, UK\\
$^{6}$Astrophysics, Department of Physics, University of Oxford, Denys Wilkinson Building, Keble Road, Oxford, OX1 3RH, UK\\
$^{7}$Centre for Radio Astronomy Techniques and Technologies, Department of Physics and Electronics, Rhodes University, PO Box 94, Makhanda 6140, South Africa\\
$^{8}$South African Radio Astronomy Observatory, 2 Fir Street, Black River Park, Observatory 7925, South Africa\\
$^{9}$Department of Physics, University of Warwick, Gibbet Hill Road, Coventry CV4 7AL, UK\\
$^{10}$School of Physics and Astronomy, University of Southampton, Highfield, Southampton SO17 1BJ, UK\\
$^{11}$International Centre for Radio Astronomy Research, Curtin University, GPO Box U1987, Perth, WA 6845, Australia\\
$^{12}$Department of Physics, University of Alberta, CCIS 4-181, Edmonton AB, T6G 2E1, Canada\\
}

\date{Accepted XXX. Received YYY; in original form ZZZ}

\pubyear{}

\begin{document}
\label{firstpage}
\pagerange{\pageref{firstpage}--\pageref{lastpage}}
\maketitle

\begin{abstract}
A program to search for radio emission from dwarf-novae-type cataclysmic variables was conducted with the South African MeerKAT radio telescope. The dwarf novae RU Pegasi, V426 Ophiuchi and IP Pegasi were detected during outburst at L-band (\qty{1284}{MHz} central frequency). Previously, only one cataclysmic variable was radio-detected at a frequency this low. We now bring the number to four. With these three newly found radio-emitters, the population of dwarf novae confirmed to be radio-emitting at any frequency reaches 10 systems. We found that the radio luminosity is correlated with the optical luminosity. For V426 Ophiuchi and RU Pegasi we found a radio decline contemporary with the outburst's optical decline. The peak radio luminosity of dwarf novae in outburst is very similar to that of novalike Cataclysmic Variables and no correlation with orbital period is seen.
\end{abstract}


%%%%%%%%%%%%%%%%%%%%%%%%%%%%%%%%%%%%%%%%%%%%%%%%%%

%%%%%%%%%%%%%%%%% BODY OF PAPER %%%%%%%%%%%%%%%%%%

\section{Introduction}
A cataclysmic variable (CV) is an interacting binary system containing an accreting white dwarf and a low-mass main sequence or low-mass giant donor star which transfers mass via Roche lobe overflow. See \citet{warnerCataclysmicVariableStars1995} for a review.

A subdivision of the systems with a main sequence donor is made based on the presence of an accretion disc.  If the white dwarf's magnetic field exceeds $\sim10^5$ Gauss, the inner disc gets truncated. These truncated disc systems, for which the white dwarf rotation is not fully tidally locked to the orbital rotation, are called intermediate polars (IPs). If the white dwarf is strongly magnetic, at roughly $B>10^7$G \citep[see][e.g. Table 6.8]{warnerCataclysmicVariableStars1995}, accretion happens along magnetic field lines and an accretion disc is not present. If the white dwarf rotation is tidally locked to the binary rotation, these systems are referred to as polars. A non-magnetic white dwarf system has an accretion disc reaching to the white dwarf surface, where the transition region is referred to as the boundary layer. 

Non-magnetic systems are further divided in dwarf novae (DNe) and novalikes (NLs). A DN possesses a disc which undergoes recurring state changes: from quiescence (low state) to outburst (high state) and back. The mechanism by which this happens is believed to be a thermal instability in the accretion disc. This idea led to the development of the so-called disc instability model (DIM, see e.g. \citet{lasotaDiscInstabilityModel2001,hameuryReviewDiscInstability2020} for a review and \citet{meyerElusiveCauseCataclysmic1981, cannizzoConvectiveAccretionDisks1982, smakAccretionCataclysmicBinaries1982,faulknerEvolutionAccretionDisc1983, mineshigeDiskinstabilityModelOutbursts1983} for the initial development of the DIM in the 1980s). In this model, the accretion discs can either be in a hot stable state (NLs), an oscillating state where the system switches from an unstable hot state with high disc accretion rate to an unstable low state with low disc accretion rate and back (DNe), or be on a stable low state branch. The outburst recurrence time-scale for DNe -- on which they switch between the unstable low state and the unstable high state and back again -- can be anything from days to decades or longer. The DNe are further divided into three subtypes (Z Cam, SU UMa and its sub-subtype WZ Sge, U Gem/SS Cyg), based on their lightcurve behaviour \citep[][section 2.1]{warnerCataclysmicVariableStars1995}. Both the orbital period and the mass transfer rate roughly correlate with subtype, having WZ Sge at the low end, then other SU UMa systems, then U Gem/SS Cyg and finally Z Cam. However, there is a large range of mass transfer rates in each group, making the mentioned grouping overlapping. The NLs as a group have even higher mass transfer rate.

NLs are closely related to DNe, but they have a sufficiently large mass transfer rate that the high state is maintained indefinitely, see e.g. \citet{dubusTestingDiskInstability2018}.

\begin{table*}
   \begin{threeparttable}
   \caption{DNe previously detected in radio}
   \label{tab:PrevDNe}
   \begin{tabular}{l l l l l l l l l}
    \hline\hline
    Object          & Orbital period        & Inclination\tnote{c}  & Paper                                                  & Telescope              & Obs. freq.                & Flux                                & Uncertainty                         & Outburst? \\
                    & (h)                   & (\degree)             &                                                        &                        & (\unit{GHz})              & (\unit{\micro\jansky})              & (\unit{\micro\jansky})              &           \\
    \hline
    SU UMa          & 1.832(1)\tnote{d}     & 44\tnote{d}           & \citet{benzFirstDetectionRadio1983}                    & Effelsberg 100m        & 4.70                      & 1300                                & 300                                 & Yes  \\
    UZ Boo          & 1.4896(7)\tnote{e}    &                       & \citet{turner12cmObservationsStellar1985}              & Arecibo interferometer & 2.38                      & 2400                                & 200                                 & Unknown \\
    TY Psc          & 1.640(1)\tnote{f}     & 18 or 55\tnote{f}     & \citet{turner12cmObservationsStellar1985}              & Arecibo interferometer & 2.38                      & 10000                               & -                                   & Unknown \\
    EM Cyg          & 6.981818(1)\tnote{g}  & 67\tnote{g}           & \citet{benzVLADetectionRadio1989}                      & VLA BnA Config.        & 4.85                      & 340                                 & 90                                  & Yes \\
    SS Cyg          & 6.603113(3)\tnote{h}  & 45-56\tnote{h}        & \citet{kordingTransientRadioJet2008}                   & VLA D Config.          & 8.50                      & 1100                                & 20                                  & Yes \\
    RX And          & 5.037432(2)\tnote{i}  & 55\tnote{i}           & \citet{coppejansDwarfNovatypeCataclysmic2016}\tnote{a} & VLA C Config.          & 10.0                      & 19.6                                & 4.4                                 & Yes \\
    YZ Cnc          & 2.0862(2)\tnote{j}    & 38\tnote{j}           & \citet{coppejansDwarfNovatypeCataclysmic2016}          & VLA C Config.          & 10.0                      & 26.8                                & 5.2                                 & Yes \\
    Z Cam           & 6.956174(5)\tnote{k}  & 57\tnote{k}           & \citet{coppejansDwarfNovatypeCataclysmic2016}          & VLA C Config.          & 10.0                      & 40.3                                & 5.2                                 & Yes \\
    SU UMa\tnote{b} & 1.832(1)\tnote{d}     & 44\tnote{d}           & \citet{coppejansDwarfNovatypeCataclysmic2016}          & VLA C/B Config.        & 10.0                      & 58.1                                & 5.7                                 & Yes \\
    U Gem           & 4.2457486(1)\tnote{l} & 70\tnote{l}           & \citet{coppejansDwarfNovatypeCataclysmic2016}          & VLA B Config.          & 10.0                      & 12.7                                & 2.8                                 & Yes \\
    \hline \hline
  \end{tabular}
  \begin{tablenotes}
      \item[a] In this paper three observations were reported per object. The detection with highest flux is listed in this table.
      \item[b] This is the second detection of SU UMa. It is SU UMa's first radio detection with an interferometer.
      \item[c] Inclinations listed here are rough estimates. Details can be found in the referenced source articles.
      \item[d] Period and inclination from \citet{thorstensenSpectroscopicStudyCataclysmic1986}.
      \item[e] Superhump period, from \citet{katoSurveyPeriodVariations2014}.
      \item[f] Orbital period from \citet{thorstensenOrbitalPeriodsSeven1996}. Inclination estimates (conflicting) are from \citet{nadalinAccretionDiskWhite2001} and from \citet{szkodyInfraredLightCurves1988}.
      \item[g] Period from \citet{csizmadiaEMCygniStudy2008}. We converted the listed "Julian Heliocentric Ephemeris Date" (HJED) period in days to hours by multiplying by 24. Inclination from \citet{northMysterySolvedMass2000}.
      \item[h] Orbital period from \citet{friend8190ASodiumDoublet1990}. Inclination from \citet{bitnerMassesEvolutionaryState2007}.
      \item[i] Period from \citet{kaitchuckOrbitalPeriodRX1989}. Inclination from \citet{pattersonDistancesAbsoluteMagnitudes2011} table 1, which cites \citet{shafterMassesWhiteDwarfs1983}.
      \item[j] Period from \citet{vanparadijs60nightCampaignDwarf1994}. Inclination from \citet{shafterTimeResolvedSpectroscopicStudy1988}.
      \item[k] Period from \citet{thorstensenImprovedEphemerisCamelopardalis1995}. Inclination from \citet{ritterCatalogueCataclysmicBinaries2003}, which cites \citet{shafterMassesWhiteDwarfs1983}.
      \item[l] Period from \citet{marshDopplerImagingDwarf1990}. Inclination from \citet{zhangEclipsesCataclysmicVariables1987}.
  \end{tablenotes}
  \end{threeparttable}
\end{table*}

Radio emission from non-magnetic CVs has been reported for a limited number of sources. See Table \ref{tab:PrevDNe} for a list of all previously radio-detected DNe. The radio flux is usually of the order of a few tens of \unit{\micro\jansky} even for the nearest sources, requiring the most powerful radio telescopes for a significant detection. For DNe the radio emission is proposed to be associated with the optical outbursts, making a detection dependent on the timing of the observations. Prior to 2008, few radio detections were published (Table \ref{tab:PrevDNe}). Three of these detections were done using a single dish telescope and a two-dish interferometer. In \citet{turner12cmObservationsStellar1985} the possibility of source confusion is mentioned. \citet{benzVLADetectionRadio1989} report that for at least one of the DNe for which \citet{turner12cmObservationsStellar1985} reports a detection (UZ Boo) source confusion is indeed very likely. It also casts doubt on the detection of TY Psc. Attempts to reproduce these radio detections were not successful \citep[e.g.][]{benzVLADetectionRadio1989}.
In 2008 SS Cygni was found as a (new) radio emitting DN \citep{kordingTransientRadioJet2008}. Since then, SS Cyg has been detected many times, including once at L-band, and is now known to produce bright radio emission during outbursts \citep{miller-jonesInvestigatingAccretionDisk2010,miller-jonesAccurateGeometricDistance2013,russellReproducibleRadioOutbursts2016,mooleyRapidRadioFlaring2017,fenderLateoutburstRadioFlaring2019}.

Following the 2008 breakthrough, a survey using the Karl G. Jansky Very Large Array radio telescope (the VLA) of both NL and DN CVs \citep{coppejansNovalikeCataclysmicVariables2015, coppejansDwarfNovatypeCataclysmic2016} showed that radio emission from CVs at 10 GHz is common but faint. Five out of five observed DNe were detected in that study. More evidence that the emission is indeed faint comes from a recent study by \citet{pelisoliSurveyRadioEmission2024}, in which they surveyed VLA data, specifically 3 GHz radio continuum data from the Very Large Array Sky Survey (VLASS) Epoch 1 Quick Look Catalogue, for radio emission from white dwarfs, at an upper limit of approximately \qty{1}{\milli\jansky}. Only one possible white dwarf radio source (not a CV) was found, out of 846 000 checked white dwarfs.
Although the occurrence of radio emission associated with DN outbursts is now established, the physical origin is less clear.
The outburst behaviour of DNe in the radio bands appears similar to that of X-ray binaries (XRBs), which are accreting neutron stars and black holes. This similarity has been used to argue that the DN radio emission is due to synchrotron emission from a collimated outflow: a jet \citep{kordingTransientRadioJet2008}. See \citet{coppejansCaseJetsCataclysmic2020} for a review. For XRBs the radio emission has been shown to originate from their jets, making it possible to examine their jet properties \citep{fenderUnifiedModelBlack2004}. This suggests that also for non-magnetic CVs one can study jet properties via radio observations. This would open the possibility to use the nearby, numerous, comparatively predictable and non-relativistic DNe as a laboratory for accretion and jet physics.

However, for NL CVs and for magnetic CVs there are strong indications that at least part of the radio emission does not come from synchrotron emission, as some flaring emission was shown to be strongly circularly polarised \citep{coppejansCaseJetsCataclysmic2020, barrettRadioObservationsMagnetic2020}. Other explanations for radio emission from CVs include electron cyclotron maser emission (ECME) \citep{melroseElectroncyclotronMasersSource1982} for short duration emission and gyrosynchrotron emission \citep[e.g.][]{chanmugamRadioEmissionCataclysmic1987} for longer duration. Emission is proposed to come from near the white dwarf or from near the secondary star or from the accretion stream \citep{kurbatovPossibleMechanismRadio2019}. Besides the open question of the origin of the radio emission, the high variability of the radio emission of CVs makes it unclear which empirical parameters govern the radio loudness of the sources. It is thus crucial to have a large enough sample of well studied (outbursts of) DNe.

We undertook a MeerKAT \citep{jonasMeerKATRadioTelescope2018} survey of nearby DNe as part of the ThunderKAT large survey project \citep{fenderThunderKATMeerKATLarge2016} to expand the sample of DNe showing radio emission during outburst, and to gain deeper insight into the predominant radio emission mechanism of both DNe in outburst and NLs \citep{hewittMeerKATSurveyNearby2020}. Observations were done in L-band (1.28 GHz, with a bandwidth of 856 MHz). Here we report on three DNe, located within \qty{300}{\parsec}, showing repeatable radio emission across different outbursts in the L-band: IP Pegasi, V426 Ophiuchi and RU Pegasi. In a separate paper (Kersten et al., in preparation) we will report and discuss the various significant non-detections.

\subsection{The dwarf novae IP Peg, V426 Oph and RU Peg}

\begin{table*}
\begin{threeparttable}
    \caption{Object properties}
    \label{tab:properties}
    \begin{tabular*}{\textwidth}{l l l l l l l}
    \hline\hline
    Object name       & R.A. (hms)\tnote{a} & Distance (\unit{\parsec})\tnote{b} & Accretor Mass (\unit{M_\odot}) & Separation (\unit{R_\odot}) & Type           \\
    Gaia DR3 ID         & Decl. (dms)  & Orbital period (\unit{h}) & Donor mass (\unit{M_\odot}) & Inclination (\degree) & Donor spectral type\tnote{c} \\
    \hline\hline
    IP Peg\tnote{d}   &  23:23:08.467       & $140.15^{+0.79}_{-0.81}$           & $1.16 \pm 0.02$                & $1.472 \pm 0.009$           & U Gem / SS Cyg \\
    2824150286583562496 & +18:24:58.67 & 3.796946470(7)            & $0.55 \pm 0.02$             & $83.8 \pm 0.5$        & M4 (3210 K)                    \\
    \hline
    V426 Oph\tnote{e} &  18:07:51.692       & $190.01^{+0.63}_{-0.72}$           & $0.90 \pm 0.19$                & $2.16 \pm 0.14$             & Z Cam          \\
    4471872295941149056 & +05:51:47.21 & $6.8472 \pm 0.0023$       & $0.70 \pm 0.14$             & $59 \pm 6$        & K5V (4440 K)                    \\
    \hline
    RU Peg\tnote{f}   &  22:14:02.545       & $271.30^{+1.48}_{-1.54}$           & $1.06 \pm 0.04$                & $2.76 \pm 0.06$             & U Gem / SS Cyg \\
    2727974767550030080 & +12:42:11.33 & $8.9904 \pm 0.0048$       & $0.96 \pm 0.08$             & $43 \pm 5$     & K5V (4440 K)                    \\
    \hline
    \end{tabular*}
    \begin{tablenotes}[flushleft, para]
    \footnotesize
    \item[a] Positions are from Gaia DR3 (ICRS, epoch 2016.0) \citep{prustiGaiaMission2016, vallenariGaiaDataRelease2023}.
    \item{b} Distance from \citet{bailer-jonesEstimatingDistancesParallaxes2021} based on Gaia EDR3.
    \item[c] In parenthesis is a temperature in Kelvin corresponding to the spectral type.
    \item[d] Orbital period, masses, separation and inclination from \citet{copperwheatPhysicalPropertiesIP2010}. The ephemeris was used to get the period and its uncertainty. Donor spectral type from \citet{martinSpectroscopyPhotometryIP1987, GrootPhD1999}.
    \item[e] Orbital period, masses, separation and inclination from \citet{hessmanTimeresolvedSpectroscopyCataclysmic1988}. Donor spectral type from \citet{northSystemicVelocitiesFour2002}.
    \item[f] Orbital period from \citet{stoverSpectroscopicStudyRadial1981}. Masses, inclination and donor spectral type from \citet{dunfordRocheTomographyCataclysmic2012}. Separation was calculated using the listed masses and period.
    \end{tablenotes}
\end{threeparttable}
\end{table*}

IP Peg is a DN of the U Gem / SS Cyg subtype with an orbital period of 3.8 hrs. It was discovered in 1981 \citep{lipovetskyNewVariableStellar1981} and has been extensively studied in the optical band \citep[e.g.][]{marshSpectroscopicStudyDeeply1988, GrootPhD1999,copperwheatPhysicalPropertiesIP2010, hanSpectroscopicPropertiesDwarf2020}. 
It shows deep eclipses, of about 5 magnitudes, which made it possible to determine the physical properties accurately, as can be seen in \citet{copperwheatPhysicalPropertiesIP2010}. For the physical parameters, see also the overview in Table \ref{tab:properties}.
As can be seen in data from the American Association of Variable Star Observers (AAVSO)\footnote{https://www.aavso.org}, the system has a DN outburst once every $\sim 4$ months, usually lasting for 7 to 10 days.

The second detected object is V426 Oph. This dwarf nova is classified as subtype Z Cam, which indicates it sometimes has longer periods (`standstills') in the high state, after which it returns to the low state and resumes DN-type outbursts. It has an orbital period of almost 7 hours \citep[][see also Table \ref{tab:properties}]{hessmanTimeresolvedSpectroscopyCataclysmic1988}. There is controversy if this object should be reclassified as an intermediate polar. See for example \citet{hellierV426OphiuchiReally1990, ramsayDefiningCharacteristicsIntermediate2008}. No X-ray modulation with the WD spin frequency (the main observational property used to determine if a CV is an IP) is found, but \citet{ramsayDefiningCharacteristicsIntermediate2008} argue that this is possible if the orbital axis is aligned precisely with the WD spin axis. An absorbed X-ray spectrum supports the classification as an IP.

RU Peg is a DN of the subtype U Gem / SS Cyg, similar to IP Peg. \cite{stoverSpectroscopicStudyRadial1981} determined its orbital period of almost 9 hours through radial velocity variations. The component masses, inclination and spectral type of the donor star were determined by \citet{dunfordRocheTomographyCataclysmic2012}. See also Table \ref{tab:properties}.
In \cite{dunfordRocheTomographyCataclysmic2012} it was also shown that during outburst there was a high level of irradiation of the donor. In particular, when RU Peg was observed near the optical outburst's peak, the side facing the primary showed a large star spot.

\section{Methods}

\subsection{Selection of nearby DNe}

As a population of semi-detached binaries, CVs are abundant throughout the Milky Way \citep{inightCatalogueCataclysmicVariables2023}. Those CVs that show optical outbursts with amplitudes of a few magnitudes are found in great numbers through historic and ongoing optical transient surveys. There are currently thousands of suspected DNe, based on their outburst lightcurves\citep{ritterCatalogueCataclysmicBinaries2003, samusGeneralCatalogueVariable2017}. For the purpose of the MeerKAT survey of DNe we have restricted ourselves to the nearest DNe (within \qty{300}{\parsec}), visible from the MeerKAT observing site (source declinations, $\delta$< +20\degree). The aim was to be as complete as possible for the nearest DNe (within \qty{150}{\parsec}) and sample a representative number of different classes of dwarf novae (e.g., SU UMa, U Gem, Z Cam and WZ Sge type) out to distances of a few hundred parsec.

\citet{palaVolumelimitedSampleCataclysmic2020} presented a volume-limited sample of 42 CVs within \qty{150}{\parsec} selected from Gaia DR2 \citep{prustiGaiaMission2016, brownGaiaDataRelease2018}, which we will call `the Pala sample'. This sample is $77 \pm 10$ percent complete and contains 25 DNe: 12 WZ Sge-subtype, 10 SU UMa-type and 3 U Gem-type systems. Only two of the DNe in the Pala sample - both U Gem systems - have previously been detected as radio sources, namely U Gem \citep{coppejansDwarfNovatypeCataclysmic2016} and SS Cyg \citep[][and references therein]{kordingTransientRadioJet2008, fenderLateoutburstRadioFlaring2019}, see also Table \ref{tab:PrevDNe}.
Restricting the Pala sample to dwarf novae south of declination $+20^{\circ}$ leaves 16 dwarf novae (6 WZ Sge, 8 SU UMa and 2 U Gem systems). Over the period of our MeerKAT observations (July 2018 - September 2023) none of the WZ Sge systems south of declination $+20^{\circ}$ underwent a dwarf nova outburst. In fact only one WZ Sge system in the entire Pala sample went into outburst, namely V627 Peg (at declination $+26^{\circ}$).
Since there were not enough optical triggers for the Pala sample, we have also observed sources picked up from the monitoring of dwarf novae in outburst via the VSNET server and sources from the Ritter-Kolb database \citep{ritterCatalogueCataclysmicBinaries2003} version 7.24, released on 31 December 2015.
Even though the aim was a well selected sample, the lack of triggers lead to a partly ad-hoc selection and thus to an inhomogeneous sample.

Based on the object's declination, outburst occurrence frequency, earlier radio observations and availability of the radio telescope it was decided if a radio observation would be requested for a given optical outburst. In total we observed 12 DNe in radio. Three objects were detected, prompting a number of follow-up observations, including observations during optical quiescence. The non-detections will be discussed in a planned follow-up paper.

\subsection{Optical data}
Optical monitoring was used to find outbursts, where an outburst was taken to be present if the magnitude reached the threshold of 1 mag brighter than the quiescence level. Optical data also provided information on where in the outburst cycle we obtained our MeerKAT observations. Furthermore, such data can in principle provide information on any relation between features in radio and optical. Optical data were acquired through VSNET\footnote{vsnet-outburst@kusastro.kyoto-u.ac.jp}, from the AAVSO website, from the Asteroid Terrestrial-impact Last Alert System (ATLAS) \citep{tonryATLASHighcadenceAllsky2018}, from the Zwicky Transient Facility (ZTF) \citep{grahamZwickyTransientFacility2019} and from the All-Sky Automated Survey for Supernovae (ASAS-SN) \citep{shappeeManCurtainXRays2014, kochanekAllSkyAutomatedSurvey2017}. We corrected for Galactic extinction, since one of the detected objects is not far from the Galactic Plane. To determine extinction, we used the 3D dustmap by Leike et al. \citep{leikeResolvingNearbyDust2020}. This dustmap has a focus on nearby dust (within \qty{\sim 400}{pc}). It reports extinction in Gaia G-band in e-folds per parsec. We integrated and converted to a magnitude value. We then calculated extinction at \qty{541.4}{nm}, the so-called $a_0$. This was done by assuming that the G-band extinction is a monochromatic extinction at the filter's pivot wavelength (\qty{621.759}{nm}), and applying a correction based on an extinction curve. As extinction curve, we used one of the models from \citet{gordonOneRelationAll2023}, namely G23 with R(V) = 3.1. Next, we used the same extinction curve to calculate the extinction at the pivot wavelengths of the relevant optical filters, and used these values as correction for the measurements.

\subsection{Radio data}
Radio observations were obtained with the South African MeerKAT\footnote{Operated by the South African Radio Astronomy Observatory (SARAO).} radio telescope \citep{jonasMeerKATRadioTelescope2018}, using its L-band (856–1712 MHz) receivers. Observations were done in the wideband coarse (4k) mode or in the wideband fine (32k) mode, which means that the bandwidth was divided in 4096 or 32768 equally sized channels. During analysis we immediately averaged to 1024 channels. The integration time used was 8 seconds. Following standard procedure, a primary calibrator, used for flux and bandpass calibration, was observed first, followed by alternating between a secondary (gain) calibrator and the target field. Although full polarization information was recorded (XX, YY, XY and YX), we did not include a polarization angle calibrator. We used J1939--6342 as primary calibrator for all observations. As secondary calibrator we used J2253+1608 for IP Peg, J1733--1304 for V426 Oph and J2232+1143 for RU Peg.

\subsection{MeerKAT data analysis}
The observations have been analysed at the IDIA/Ilifu cluster\footnote{https://www.idia.ac.za/ilifu-research-cloud-infrastructure/} with the Oxkat set of python scripts \citep{heywoodOxkatSemiautomatedImaging2020}. Oxkat streamlines the process of flagging, calibration, self-calibration and direction-dependent calibration. An extensive set of underlying software is used: CASA \citep{mcmullinCASAArchitectureApplications2007, teamCASACommonAstronomy2022}, CubiCal \citep{kenyonCUBICALFastRadio2018}, DDFacet \citep{tasseFacetingDirectiondependentSpectral2018}, KillMS \citep{tasseKillMS2022}, Owlcat \citep{smirnovOwlcat2022}, Ragavi \citep{rhodesuniversitycentreforradioastronomytechniquesandtechnologiesrattRagavi2022}, ShadeMS \citep{smirnovShadeMS2022}, Singularity \citep{kurtzerSingularityScientificContainers2017, kurtzerSingularityApptainer2021, singularitycedevelopersSingularityCE2021}, Stimela \citep{makhathini2018, makhathiniStimela2022}, Tricolour \citep{hugoTricolourOptimizedSumThreshold2022, rhodesuniversitycentreforradioastronomytechniquesandtechnologiesrattTricolour2022}, WSClean \citep{offringa-wsclean-2014, offringa-wsclean-2017, offringaWSClean}. Processing by Oxkat is done in multiple steps, named INFO, 1GC, FLAG, 2GC and 3GC.

\subsubsection{Flagging}
During the 1GC step of processing the primary calibrator (used for flux density calibration and bandpass determination) scan(s) is/are flagged, using CASA autoflagging with the {\tt flagdata} task. Here the {\tt rflag, tfcrop} and {\tt extend} algorithms are used. Then static basic flagging is applied to all data: known bad frequencies are blacklisted. These include frequencies used by certain satellites. The {\tt rflag} algorithm calculates an RMS value of the polarization power per short time interval per channel (we averaged to 1024 channels) and flags the interval-channel pair if this RMS is above a certain threshold. Also, a sliding window median value is calculated for the real and for the imaginary part of the visibilities over the spectral channels at fixed time interval. If the median changes too quickly the interval-channel pair is flagged. The {\tt tfcrop} algorithm calculates a time-averaged bandpass with a polynomial fit, which does not include spikes. Then the non-averaged bandpass per time bin is compared to this one and time-frequency pairs which contain an outlier are removed. Then this procedure is repeated with the role of time bin and frequency bin swapped. A similar procedure is done for flagging the gain calibrator field and the science field. However, in these cases Tricolour is used for autoflagging instead of CASA. In some cases some extra manual flagging was done. In the end, typically a large fraction of the recorded visibilities gets flagged. For example, for V426 Oph 2022 Jul epoch 2 50.26 percent was flagged.

\subsubsection{Calibration}
Calibration is done in three `generations': 1GC, 2GC and 3GC. 1GC (first generation calibration) is calibration from the calibrator objects. During this step the reference antenna (refant) is chosen from a supplied pool. While the SARAO pipeline often uses an antenna far from the centre of the array, we always used an antenna near the centre, following the Oxkat default. Oxkat takes the initially least flagged antenna from a pool of antennas. After the 1GC calibration, performed with the calibrator data, 2GC follows. 2GC is self-calibration, based on closure phases in the target image itself. 3GC (third generation calibration level) is direction-dependent self-calibration. In the 3GC step, peeling is used to remedy rippling effects from the nearest bright source. We obtained self-calibration for all observations (2GC). Peeling (3GC) was also done for all observations. Another direction-dependent (3GC) calibration technique is faceting \citep{tasseFacetingDirectiondependentSpectral2018}. For one observation of V426 Oph (2021 May Epoch 1), faceting was performed. Science image quality improved slightly, but not enough to warrant processing the other observations with this technique.

\subsubsection{Imaging}
\label{Imaging}
Oxkat generates images with WSClean, at the 1CG level and at the 2GC level. It does not generate an image after the 3GC peeling step. So, after finishing calibration we did manual imaging directly with WSClean, outside of Oxkat. CARTA, the Cube Analysis and Rendering Tool for Astronomy \citep{comrieCARTACubeAnalysis2021}, was then used to inspect the resulting science images and to determine an RMS value\footnote{RMS: Root Mean Square. In the context of flux density considerations, it is the root of the mean of the square of the per-pixel deviation from the mean value in an area of the radio image.} around the target source. Flux density of the radio sources has been determined with PyBDSF. See Section \ref{FluxDetermination}.
For imaging we used Briggs robust weighting of --0.3 for V426 Oph and RU Peg, and --0.05 for IP Peg. Compared to Oxkat's prepeel image, our images are more naturally weighted, which improves flux density measurement at the expense of a decrease in spatial resolution in the image. For V426 Oph we used the local RMS option of WSClean, although with larger than default local-rms-window value (200 instead of 25).

\subsubsection{Flux density and RMS determination}
\label{FluxDetermination}
We determined the pixel RMS value in the neighbourhood of the source that we are interested in. We do this with the science image created with WSClean. For each source we define a region (usually a rectangle or circle) very near the source's Gaia DR3 position, of at least 1000 pixels, but usually a lot more. We avoided any obviously synthesized beam shaped blobs. In this region we measure, using CARTA, the pixel RMS value. This value is reported as the RMS for upper limits. Flux density has been determined with PyBDSF \citep{pybdsfprogrammersPyBDSF2022}. We used the 1.10.4 development version. 
The calibrated science image of the observed field is fed as input into PyBDSF. The program uses a moving square (we used 80x80 pixels, except near bright sources, where we changed to 20x20 pixels) to find the local RMS. The RMS found near the target is close to the one determined manually, but sometimes slightly higher, probably because sources below 5 sigma are present in these L-band images. We report the island RMS (RMS in a region determined during source detection, with all its points very near the source for our observations) in Table \ref{tab:luminosities}.

As a check, we also determined the flux density for some observations with CARTA, using the FluxDensity output in the statistics widget, and using a beam-sized region. These flux densities showed good agreement with the PyBDSF flux density values.

In our L-band images calibration is not able to completely remove artefacts from bright sources. There is a spike pattern going out from these sources, which rotates with time. This can be a source of uncertainty which is not completely reflected in the reported local RMS value.

For a number of radio sources we compared PyBDSF-reported flux densities between the different observations of the same field. We looked at sources with a flux density of at least \qty{300}{\micro\jansky}. These sources are not the target source, and are assumed to have constant radio flux density. However, the flux density for each source found in PyBDSF can vary by 5 to 10 percent between epochs, which is usually a bit more than the PyBDSF-reported uncertainty would indicate. This suggests that neither the reported RMS nor the PyBDSF-reported uncertainty captures all sources of uncertainty, and the real uncertainty is higher. Bright source artefacts depend on the time of observation. But even with the same observation time, slightly different calibration, imaging and measurement settings will result in uncertainty up to this 10 percent level based on artefacts near the measured source. Uncertainty of the absolute flux calibration will also be a cause of this variation.

\subsubsection{Distance and specific luminosity determination}
For each detection the specific radio luminosity\footnote{Radio luminosity per unit of frequency, measured in \unit{W.Hz^{-1}} or \unit{erg.s^{-1}.Hz^{-1}}, is sometimes simply called radio luminosity, but more properly called spectral radio luminosity. The term specific radio luminosity is also in use. In this article we follow this convention: we use the term specific radio luminosity.} is determined based on the measured flux density and the distance. We used the distance estimates of \citet{bailer-jonesEstimatingDistancesParallaxes2021}, based on Gaia EDR3 data \citep{brownGaiaEarlyData2021, brownGaiaEarlyData2021a}. The distance uncertainty is not Gaussian (see Table \ref{tab:properties}). For calculating the specific luminosity uncertainty we approximate the distance uncertainty by taking the largest of the lower and upper bound as the sigma of a Gaussian distribution. For non-detections a specific luminosity upper limit is calculated. We use 3 times the RMS (as measured in CARTA) as the upper limit estimate for the flux density.

\subsection{Radio observations}

\begin{table*}
   \begin{threeparttable}
   \caption{MeerKAT observations of IP Peg, V426 Oph and RU Peg}
   \label{tab:obs}
   \begin{tabular}{l l l l l l l}
    \hline\hline
    Object & Observation ID \tnote{a} & Start time (UT) & End time (UT) & Mid Time (MJD) & TOS (h) \tnote{b} & Antennas \\
    \hline
    IP Peg & 2021\,Jun epoch 1        & 2021-06-04 04:57:03 & 2021-06-04 06:11:17 & 59369.23206 & 1 & 62 \\
    IP Peg & 2021\,Jun quiescence     & 2021-06-20 00:44:43 & 2021-06-20 01:58:33 & 59385.05669 & 1 & 60 \\
    IP Peg & 2021\,Sep epoch 1        & 2021-09-20 19:49:12 & 2021-09-21 00:21:05 & 59477.92023 & 4 & 57 \\
    IP Peg & 2021\,Sep epoch 2        & 2021-09-21 18:09:03 & 2021-09-21 22:42:40 & 59478.85130 & 4 & 57 \\
    IP Peg & 2022\,Jan epoch 1        & 2022-01-10 12:33:05 & 2022-01-10 17:10:02 & 59589.61913 & 4 & 61 \\
    V426 Oph & 2021\,May epoch 1      & 2021-05-14 02:18:07 & 2021-05-14 03:33:57 & 59348.12224 & 1 & 61 \\
    V426 Oph & 2021\,Jun quiescence   & 2021-06-30 22:11:47 & 2021-06-30 23:27:21 & 59395.95109 & 1 & 59 \\
    V426 Oph & 2022\,Jul epoch 1      & 2022-07-05 20:33:55 & 2022-07-05 21:54:41 & 59765.88493 & 1 & 63 \\
    V426 Oph & 2022\,Jul epoch 2      & 2022-07-06 21:05:42 & 2022-07-06 22:26:19 & 59766.90695 & 1 & 63 \\
    V426 Oph & 2022\,Jul epoch 3      & 2022-07-07 21:17:47 & 2022-07-07 22:38:17 & 59767.91531 & 1 & 60 \\
    V426 Oph & 2022\,Jul epoch 4      & 2022-07-09 17:32:50 & 2022-07-09 18:52:56 & 59769.75895 & 1 & 63 \\
    V426 Oph & 2022\,Jul epoch 5      & 2022-07-11 21:57:50 & 2022-07-11 23:18:44 & 59771.94325 & 1 & 63 \\
    V426 Oph & 2022\,Jul epoch 6      & 2022-07-12 17:32:50 & 2022-07-12 18:52:56 & 59772.75895 & 1 & 63 \\
    V426 Oph & 2022\,Jul epoch 7      & 2022-07-13 20:22:50 & 2022-07-13 21:43:28 & 59773.87718 & 1 & 63 \\
    RU Peg & 2021\,Apr epoch 1        & 2021-04-27 06:18:24 & 2021-04-27 07:32:14 & 59331.28841 & 1 & 63 \\
    RU Peg & 2021\,Apr epoch 2        & 2021-04-29 06:22:24 & 2021-04-29 07:35:59 & 59333.29110 & 1 & 59 \\
    RU Peg & 2021\,May quiescence     & 2021-05-28 03:32:49 & 2021-05-28 04:46:39 & 59362.17342 & 1 & 63 \\
    RU Peg & 2022\,Jun epoch 1        & 2022-06-03 00:12:55 & 2022-06-03 02:36:51 & 59733.05894 & 2 & 63 \\
    RU Peg & 2022\,Jun epoch 2        & 2022-06-04 00:07:53 & 2022-06-04 02:31:50 & 59734.05545 & 2 & 63 \\
    RU Peg & 2022\,Jun epoch 3        & 2022-06-05 00:08:04 & 2022-06-05 02:32:00 & 59734.05545 & 2 & 63 \\
    RU Peg & 2022\,Jun epoch 4        & 2022-06-06 00:02:52 & 2022-06-06 02:26:49 & 59736.05197 & 2 & 63 \\
    RU Peg & 2022\,Jun epoch 5        & 2022-06-06 23:58:38 & 2022-06-07 02:22:26 & 59737.04897 & 2 & 59 \\
    RU Peg & 2022\,Jun epoch 6        & 2022-06-07 23:53:36 & 2022-06-08 02:17:24 & 59738.04548 & 2 & 58 \\
    RU Peg & 2022\,Jun epoch 7        & 2022-06-08 23:47:42 & 2022-06-09 02:11:46 & 59739.04147 & 2 & 59 \\
    RU Peg & 2022\,Jun epoch 8        & 2022-06-09 23:43:49 & 2022-06-10 02:07:45 & 59740.03873 & 2 & 61 \\
    \hline \hline
    \end{tabular}
    \begin{tablenotes}
    \item[a] We include an epoch number in each outburst observation name, even if only one epoch was taken during a given outburst.
    \item[b] Time on source.
    \end{tablenotes}
    \end{threeparttable}
\end{table*}

For all initial outburst observations and the observations during quiescence, the time on source was 1 hour, while most follow-up observations had a longer duration. A list of MeerKAT observations is in Table \ref{tab:obs}. The observations were obtained from April 2021 to July 2022. Between 57 and 63 dishes were active for these observations.

\section{Results}

\begin{table*}
   \hspace*{-1cm}
   \centering
   \begin{threeparttable}
   \caption{Flux densities and luminosities}
   \label{tab:luminosities}
   \begin{tabular}{l l l l l l l}
    \hline\hline
    \rule{0pt}{8.0pt}
    Object & Observation ID & Flux density (\unit{\micro\jansky}) & RMS (\unit{\micro\jansky}) & Significance & Luminosity ($10^{15}$\unit{erg.s^{-1}.Hz^{-1}}) & Separation (arcsec) \tnote{a} \\
    \hline
    IP Peg   & 2021\,Jun epoch 1      & $ 84.2 \pm 14.4$             & 11.9 &  5.9 & $ 1.98   \pm 0.34$ & 0.36 \\
    IP Peg   & 2021\,Jun quiescence   & $ 65.3 \pm 16.4$\tnote{b}    & 13.5 &  4.0 & $ 1.53 \pm 0.39$ & 1.13 \\
    IP Peg   & 2021\,Sep epoch 1      & $ 69.7 \pm 10.0$             &  8.3 &  7.0 & $ 1.64 \pm 0.24$ & 0.38 \\
    IP Peg   & 2021\,Sep epoch 2      & $ 79.6 \pm 10.3$             &  8.5 &  7.7 & $ 1.87 \pm 0.24$ & 0.74 \\
    IP Peg   & 2022\,Jan epoch 1      & $ 75.7 \pm  8.9$             &  7.3 &  8.5 & $ 1.78 \pm 0.21$ & 0.36 \\
    V426 Oph & 2021\,May epoch 1      & $598.4 \pm 34.9$             & 28.7 & 17.1 & $25.8  \pm 1.5$ & 0.51 \\
    V426 Oph & 2021\,Jun quiescence   &             $-$              & 27.6 &  $-$ & $< 1.19$ & $-$ \\
    V426 Oph & 2022\,Jul epoch 1      &             $-$              & 26.3 &  $-$ & $< 1.14$ & $-$ \\
    V426 Oph & 2022\,Jul epoch 2      & $111.7 \pm 38.4$\tnote{b}    & 31.5 &  2.9 & $ 4.8 \pm 1.7$ & 0.88 \\
    V426 Oph & 2022\,Jul epoch 3      & $118.4 \pm 40.3$\tnote{b}    & 32.9 &  2.9 & $ 5.1 \pm 1.7$ & 0.89 \\
    V426 Oph & 2022\,Jul epoch 4      & $250.9 \pm 58.4$\tnote{b}    & 47.8 &  4.3 & $ 10.8 \pm 2.5$ & 1.76 \\
    V426 Oph & 2022\,Jul epoch 5      & $174.4 \pm 34.4$             & 28.2 &  5.1 & $ 7.5 \pm 1.5$ & 1.01 \\
    V426 Oph & 2022\,Jul epoch 6      & $126.0 \pm 38.6$\tnote{b}    & 31.6 &  3.3 & $ 5.4 \pm 1.7$ & 0.48 \\
    V426 Oph & 2022\,Jul epoch 7      &             $-$              & 24.0 &  $-$ & $< 1.04$ & $-$ \\
    RU Peg   & 2021\,Apr epoch 1      & $156.8 \pm 16.6$             & 13.6 &  9.5 & $13.8  \pm 1.5 $ & 0.28 \\
    RU Peg   & 2021\,Apr epoch 2      & $136.1 \pm 15.5$             & 12.8 &  8.8 & $12.0  \pm 1.4 $ & 0.88 \\
    RU Peg   & 2021\,May quiescence   & $ 59.7 \pm 12.1$             & 10.0 &  4.9 & $ 5.26 \pm 1.07 $ & 0.55 \\
    RU Peg   & 2022\,Jun epoch 1      & $135.2 \pm 10.5$             &  8.6 & 12.9 & $11.9  \pm 0.9 $ & 0.26 \\
    RU Peg   & 2022\,Jun epoch 2      & $104.3 \pm 10.4$             &  8.5 & 10.1 & $ 9.19 \pm 0.92$ & 0.43 \\
    RU Peg   & 2022\,Jun epoch 3      &  $86.0 \pm 10.4$             &  8.5 &  8.3 & $ 7.57 \pm 0.92$ & 0.32 \\
    RU Peg   & 2022\,Jun epoch 4      &  $63.7 \pm  9.9$             &  8.1 &  6.4 & $ 5.61 \pm 0.87$ & 1.24 \\
    RU Peg   & 2022\,Jun epoch 5      &  $45.7 \pm 10.5$             &  8.6 &  4.4 & $ 4.03 \pm 0.93$ & 0.31 \\
    RU Peg   & 2022\,Jun epoch 6      &  $57.2 \pm 12.4$             & 10.2 &  4.6 & $ 5.04 \pm 1.09$ & 1.08 \\
    RU Peg   & 2022\,Jun epoch 7      &  $53.0 \pm 17.2$             & 10.0 &  3.1 & $ 4.66 \pm 1.52 $ & 2.83 \\
    RU Peg   & 2022\,Jun epoch 8      &  $55.7 \pm 11.1$             &  9.1 &  5.0 & $ 4.91 \pm 0.97 $ & 1.61 \\
   \hline \hline
   \end{tabular}
   \begin{tablenotes}
   \item[a] The angular separation between the centre of the fitted radio source and the Gaia DR3 position of the object. For comparison, a typical restoring beam width was $\sim$8 arcsec.
   \end{tablenotes}
   \end{threeparttable}
   \hspace*{-1cm}
\end{table*}

\begin{figure*}
  \resizebox{0.92\linewidth}{!}{\includegraphics{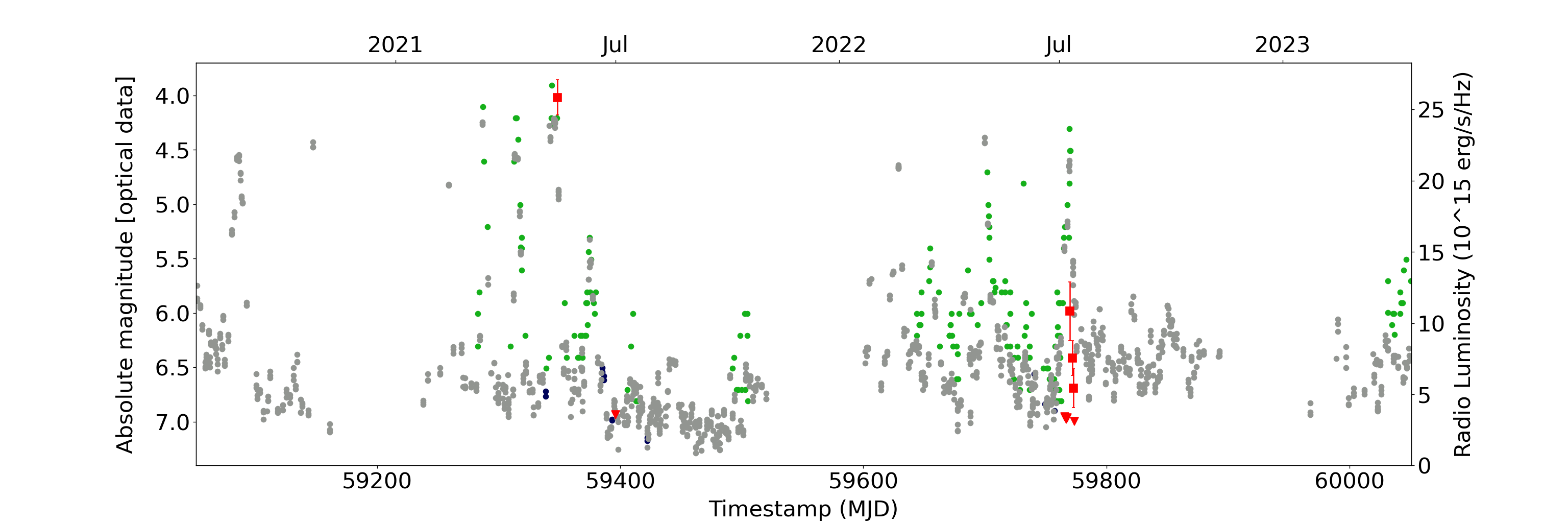}}
  \resizebox{0.92\linewidth}{!}{\includegraphics{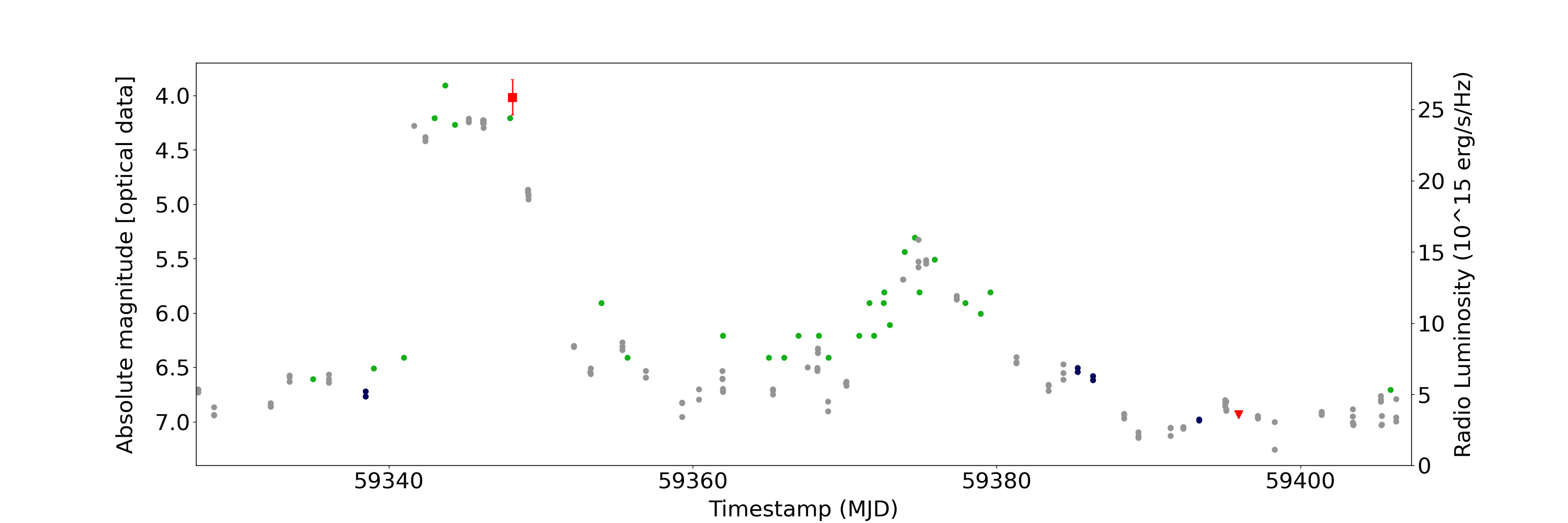}}
  \resizebox{0.92\linewidth}{!}{\includegraphics{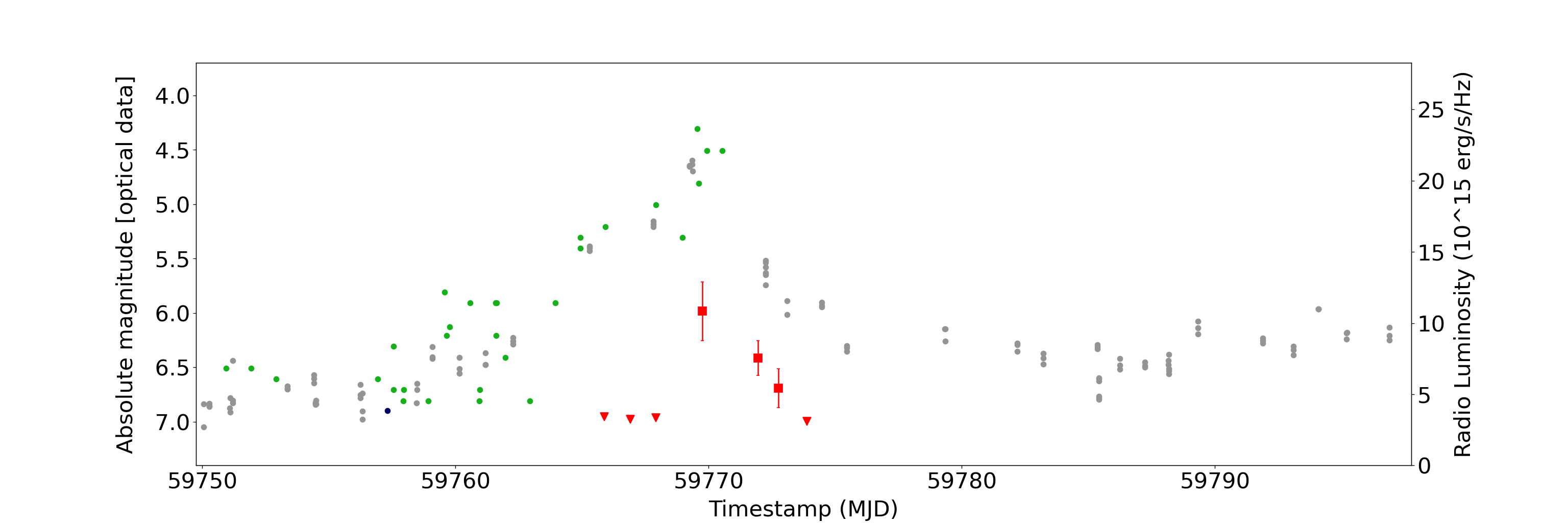}}
  \caption{Optical magnitude and radio luminosity versus time for V426 Oph. Panel a (top): 1000 days, starting on 2021-04-27. Panel b (middle) shows 80 days and panel c (bottom) shows 48 days, giving a more detailed view. The optical data in the background are from ASAS-SN g-filter (grey) and ZTF g-filter (blue), VSNET (green). Radio error bars reflect the local RMS and source fitting. Plotted radio upper limits are 3 sigma upper limits.}
  \label{fig:OpticalRadioLightcurveV426Oph}
\end{figure*}

\begin{figure*}
  \resizebox{0.92\linewidth}{!}{\includegraphics{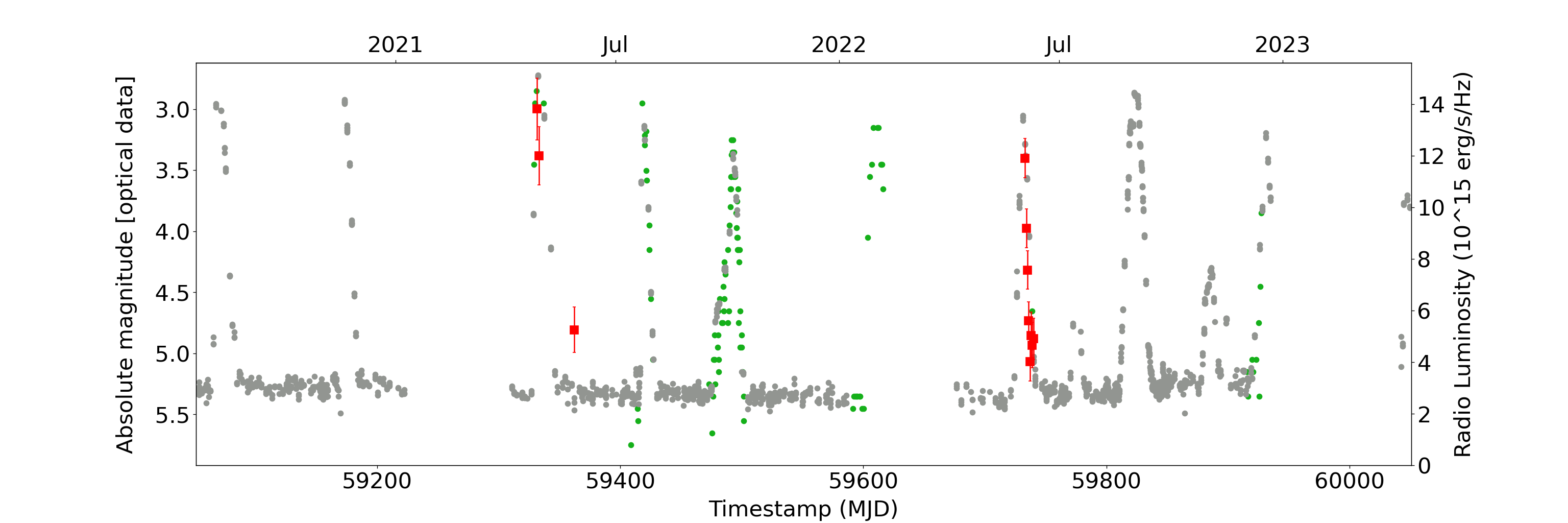}}
  \resizebox{0.92\linewidth}{!}{\includegraphics{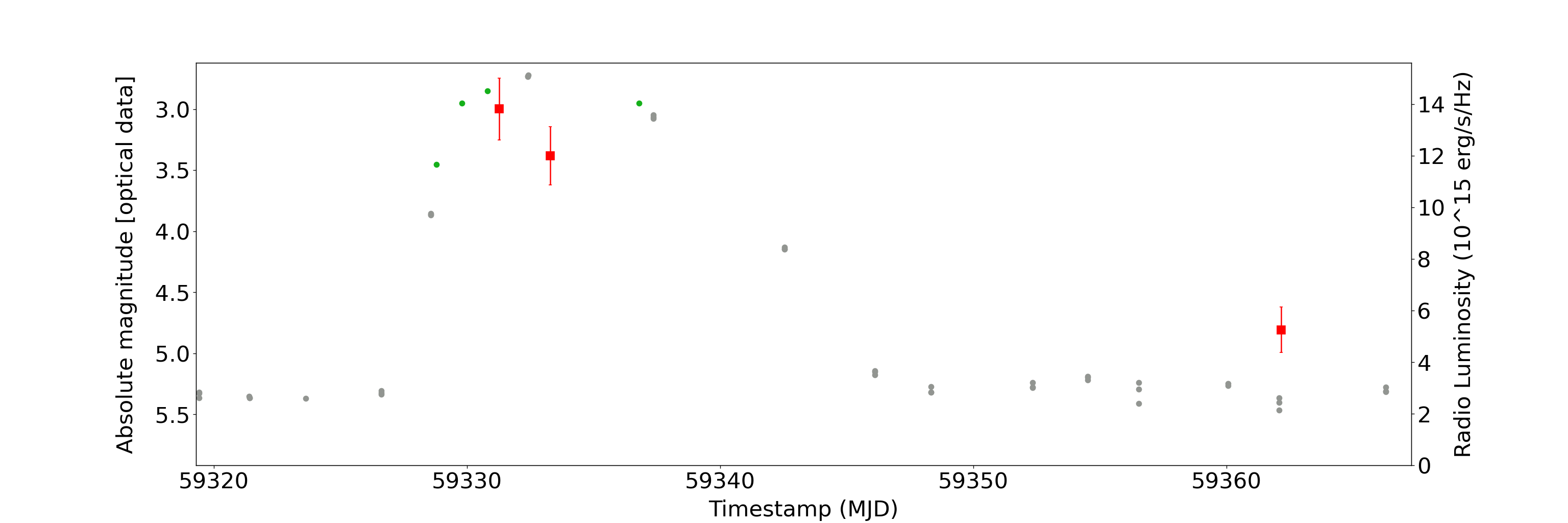}}
  \resizebox{0.92\linewidth}{!}{\includegraphics{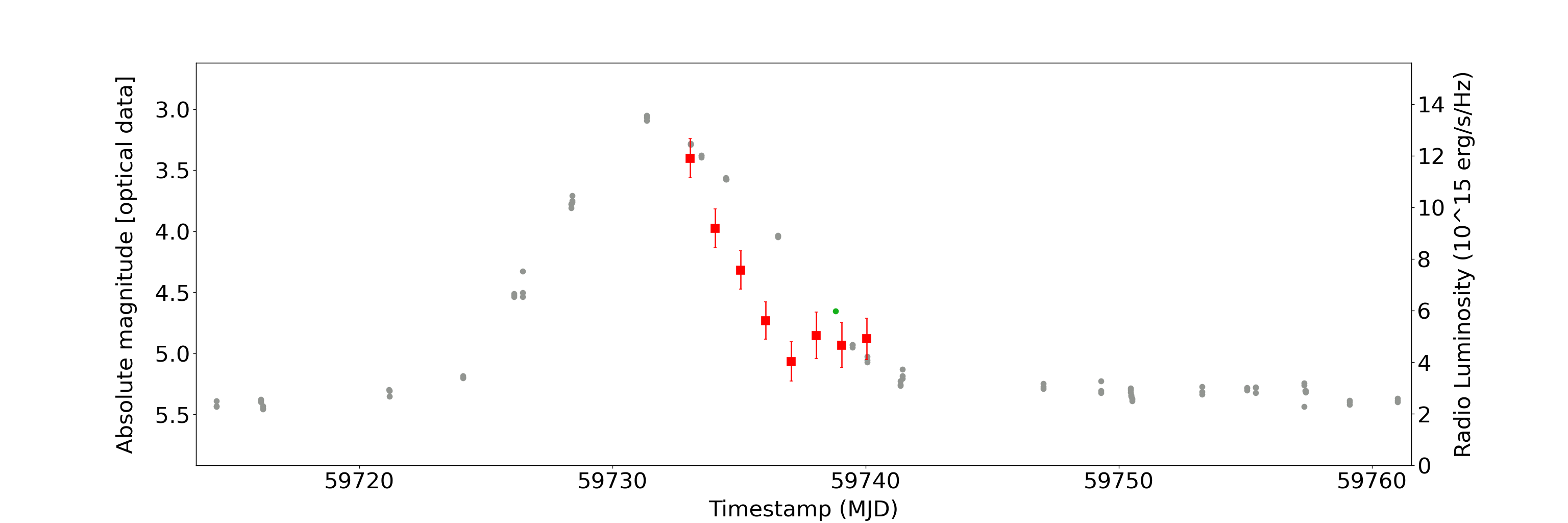}}
  \caption{Optical magnitude and radio luminosity versus time for RU Peg. Panel a (top): 1000 days, starting on 2021-04-27. The other panels, b (middle) and c (bottom), each show 48 days, giving a more detailed view. The optical data in the background are from ASAS-SN g-filter (grey), VSNET (green). Radio error bars reflect the local RMS and source fitting.}
  \label{fig:OpticalRadioLightcurveRUPeg}
\end{figure*}

Table \ref{tab:luminosities} shows the measured flux density and RMS for each observation. Figures \ref{fig:OpticalRadioLightcurveIPPeg}, \ref{fig:OpticalRadioLightcurveV426Oph} and \ref{fig:OpticalRadioLightcurveRUPeg} show the optical and radio light curves for the three objects, IP Peg, V426 Oph and RU Peg, respectively.
Optical data have been converted to absolute magnitude and radio fluxes to luminosity, using Gaia DR3 distances. The optical data were taken in multiple filters.

\subsection{IP Peg}
\begin{figure*}
  \resizebox{0.92\linewidth}{!}{\includegraphics{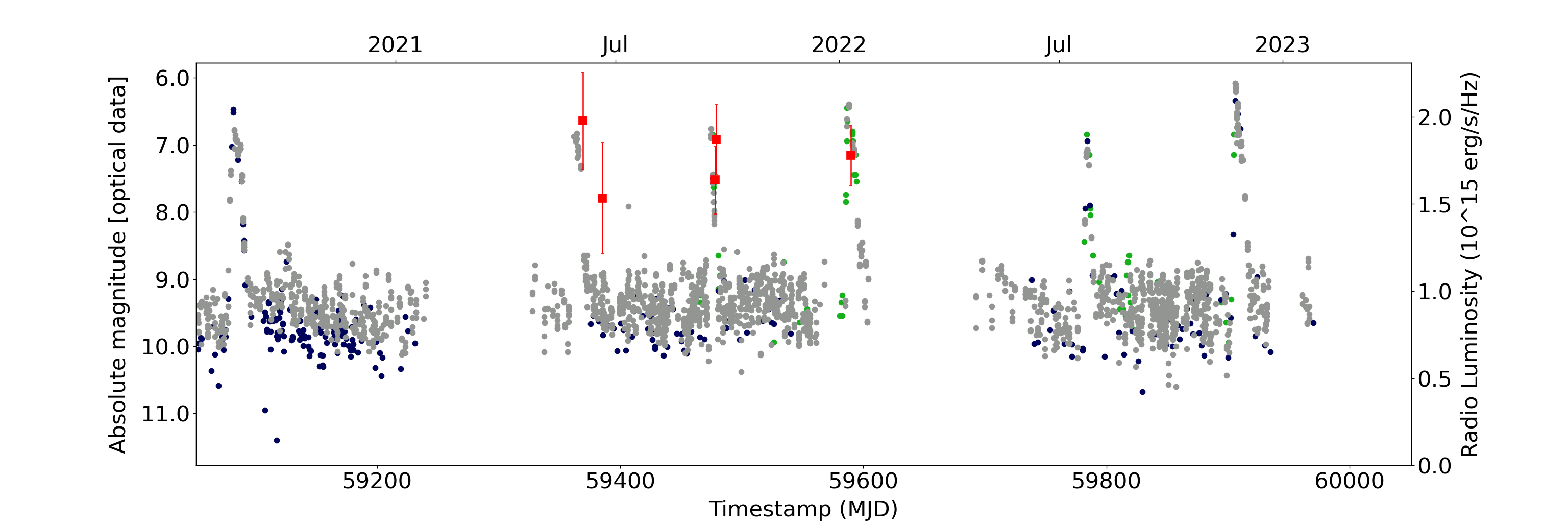}}
  \resizebox{0.92\linewidth}{!}{\includegraphics{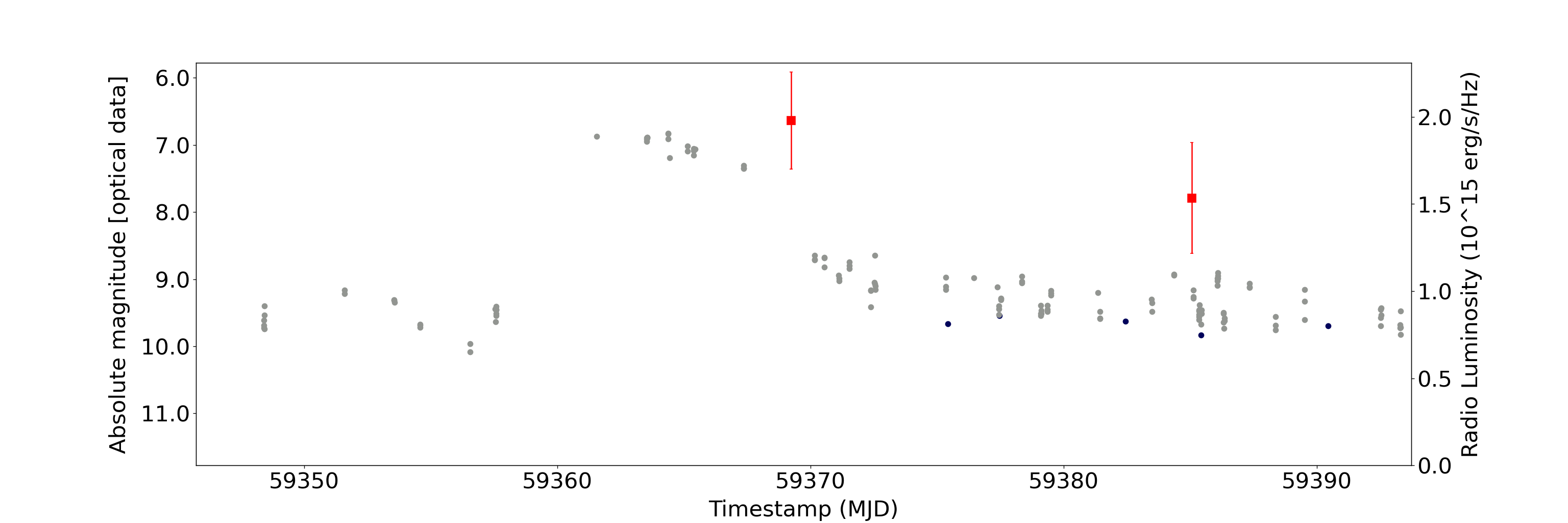}}
  \resizebox{0.92\linewidth}{!}{\includegraphics{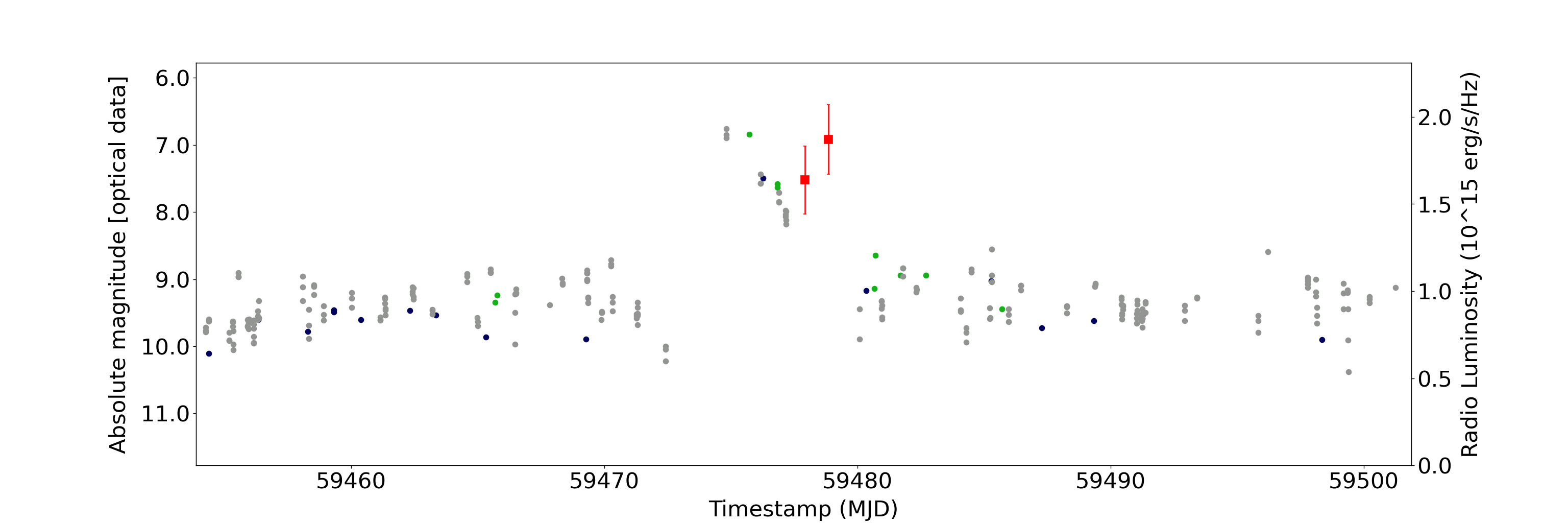}}
  \resizebox{0.92\linewidth}{!}{\includegraphics{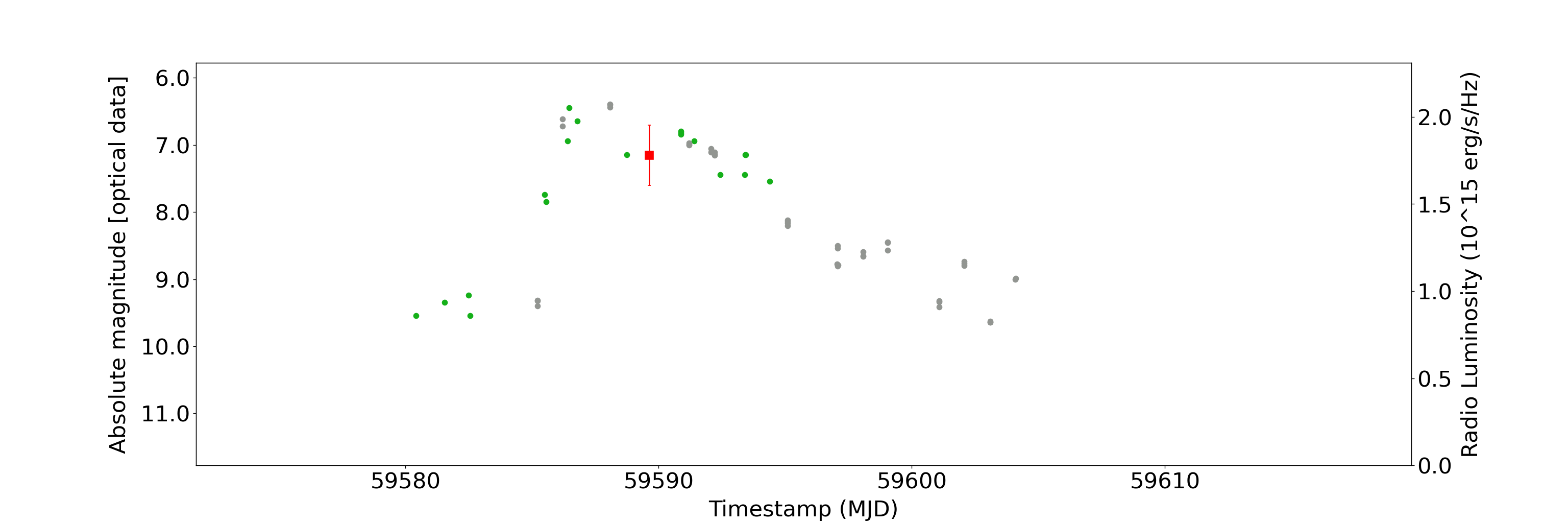}}
  \caption{Optical magnitude and radio luminosity versus time for IP Peg. Panel a (top): 1000 days, starting on 2021-04-27. The other panels (b, c, d from top to bottom) each show 48 days, giving a more detailed view. The optical data in the background are from ASAS-SN g-filter (grey), ZTF g-filter (blue), VSNET (green). Radio error bars reflect the local RMS and source fitting.}
  \label{fig:OpticalRadioLightcurveIPPeg}
\end{figure*}

In Figure \ref{fig:OpticalRadioLightcurveIPPeg} the optical data of IP Peg show multiple outbursts.
Our initial MeerKAT observation of IP Peg was during its June 2021 (MJD 59355) outburst (Figure \ref{fig:OpticalRadioLightcurveIPPeg} panel b). This outburst's duration was 10 days in the optical, with a plateau phase followed by the decline. Our radio observation occurred a few days after the optical peak was reached, when the decline was underway. The optical quiescence level would be reached a little over one day later. The outburst had its optical peak at absolute magnitude approximately 6.9 . This is a bit fainter than most of IP Peg's other outbursts, including the outburst of January 2022 (following MJD 59580) which reached absolute magnitude 6.5 (Fig. \ref{fig:OpticalRadioLightcurveIPPeg} panel d). The first radio observation was followed by a quiescence observation $\sim$15 days after the outburst (Fig. \ref{fig:OpticalRadioLightcurveIPPeg} panel b). Additional observations were obtained in September 2021 and during the January 2022 outburst. The September 2021 (MJD 59478) outburst was shorter in the optical (about 5 days long), had a quick rise, peaked, and then started the decline. Both the rise and decline rate appear almost linear in magnitude, but with different gradients.

The four radio observations of IP Peg in outburst resulted in four detections with more than 5-sigma significance. During quiescence we still formally detect the source with roughly 4-sigma, however, we needed to lower the significance thresholds to find the source in PyBDSF.
The radio luminosity is consistent across the observations taken during the three different outbursts, with flux density from \qty{70 \pm 10}{\micro\jansky} to \qty{84 \pm 14}{\micro\jansky}. The quiescence observation has consistent (but the lowest) luminosity, at flux density \qty{65 \pm 16}{\micro\jansky}. At $\sim$\qty{80}{\micro\jansky} a chance alignment of a background source is unlikely, so we interpret these observations as detection of IP Peg. Usually, CVs are not detected during quiescence and show only significant radio emission during the outburst. Given the lower significance of the quiescence flux detection, we present it with caution.

IP Peg's initial (June 2021) radio observation was obtained around the time that the optical decline started. The second observed outburst took place about three months later. Two observations were done when the optical decline had already set in. These two radio observations show consistent luminosities, at \qty{1.6e15}{erg.s^{-1}.Hz^{-1}} and \qty{1.9e15}{erg.s^{-1}.Hz^{-1}}. The third observed outburst had an observation which happened after the optical peak, but near the beginning of the optical decline. IP Peg is detected consistently during a dwarf nova outburst.

\subsection{V426 Oph}
The first and second MeerKAT observations of V426 Oph are indicated in Figure \ref{fig:OpticalRadioLightcurveV426Oph} panel b. During the initial May 2021 (following MJD 59340) outburst the radio observation was obtained during the peak/plateau, just before the optical decline started. This was a detection, at high luminosity ( \qty{2.6e16}{erg.s^{-1}.Hz^{-1}} ). No significant change in flux density during the one hour of observation was found. There is a rebrightening in the optical in the beginning of June. We obtained our quiescence observation, a non-detection, later that month, at MJD 59396, after the rebrightening finished.

Figure \ref{fig:OpticalRadioLightcurveV426Oph} panel c shows a follow-up series of radio observations in July 2022 (following MJD 59760). These observations were taken before, at, and after the optical peak. The peak optical absolute magnitude was $\sim$4.7, which is a little fainter than the 2021 May peak, which was at absolute magnitude 4.2 . Furthermore, the optical rise is much slower than for the May 2021 outburst. This indicates that the outburst heating wave travelled inside-out. See for example \citet{katoAnalysisThreeSU2013} for discussion of outside-in a and inside-out classification based on lightcurves.
During the optical rise there are radio non-detections. Right before the optical peak, radio emission is detected for the first time during this outburst, meaning that the radio rise happened many days after the optical rise started, which means many days after the disc heating wave initiated. For this outburst there is no optical plateau phase. Instead, the optical decline sets in directly after reaching peak optical luminosity. Radio emission is detected twice more during this decline. The last observation results in a non-detection with a 3 sigma upper limit. Summarizing, we see that radio emission is not found during the optical rise, but abruptly starts around the optical peak, and smoothly declines soon after its quick rise, going under the detection limit when optical quiescence is reached. The source's radio emission did not become nearly as bright as in the 2021 detection, but still resulted in a clear detection.

\subsection{RU Peg radio and optical lightcurves}
In Figure \ref{fig:OpticalRadioLightcurveRUPeg} the optical data for RU Peg show that its magnitude remains steady during quiescence phases, and outbursts are sharply visible. Figure \ref{fig:OpticalRadioLightcurveRUPeg} panel b shows the lightcurve of the outburst in April 2021 (following MJD 59320). The first two radio observations were taken near the peak of the outburst. The third observation was obtained 16 days after quiescence was reached. The quiescence radio emission drops to about a third of what it was near the peak: from \qty{13.8\pm1.5}{\times 10^{15} erg.s^{-1}.Hz^{-1}} to \qty{5.3\pm1.1}{\times 10^{15} erg.s^{-1}.Hz^{-1}}.

A series of radio observations was obtained during the June 2022 outburst (following MJD 59720). The radio observations started right after the optical peak was reached. See Figure \ref{fig:OpticalRadioLightcurveRUPeg} panel c. At the time of the first data point the radio luminosity is similar to what it was during the outburst detections in 2021, at \qty{11.9\pm0.9}{\times 10^{15} erg.s^{-1}.Hz^{-1}}. The first four radio points show the radio decline. The second set of four data points shows a luminosity which agrees with the quiescent data point found in 2021: at or below \qty{4}{\times 10^{15} erg.s^{-1}.Hz^{-1}}. This quiescent level can be emission from RU Peg, but a background source is also possible, given the density of sources with flux density of approximately \qty{50}{\micro\jansky} present in the science image. It is not clear if the first detection is near the radio peak, or if the radio decline had already set in. However, the radio luminosity declined to quiescent level before the optical outburst ended, with a roughly linear decline rate.

\section{Discussion}

\subsection{Radio luminosity versus orbital period}
\begin{figure*}
  \resizebox{\hsize}{!}{\includegraphics{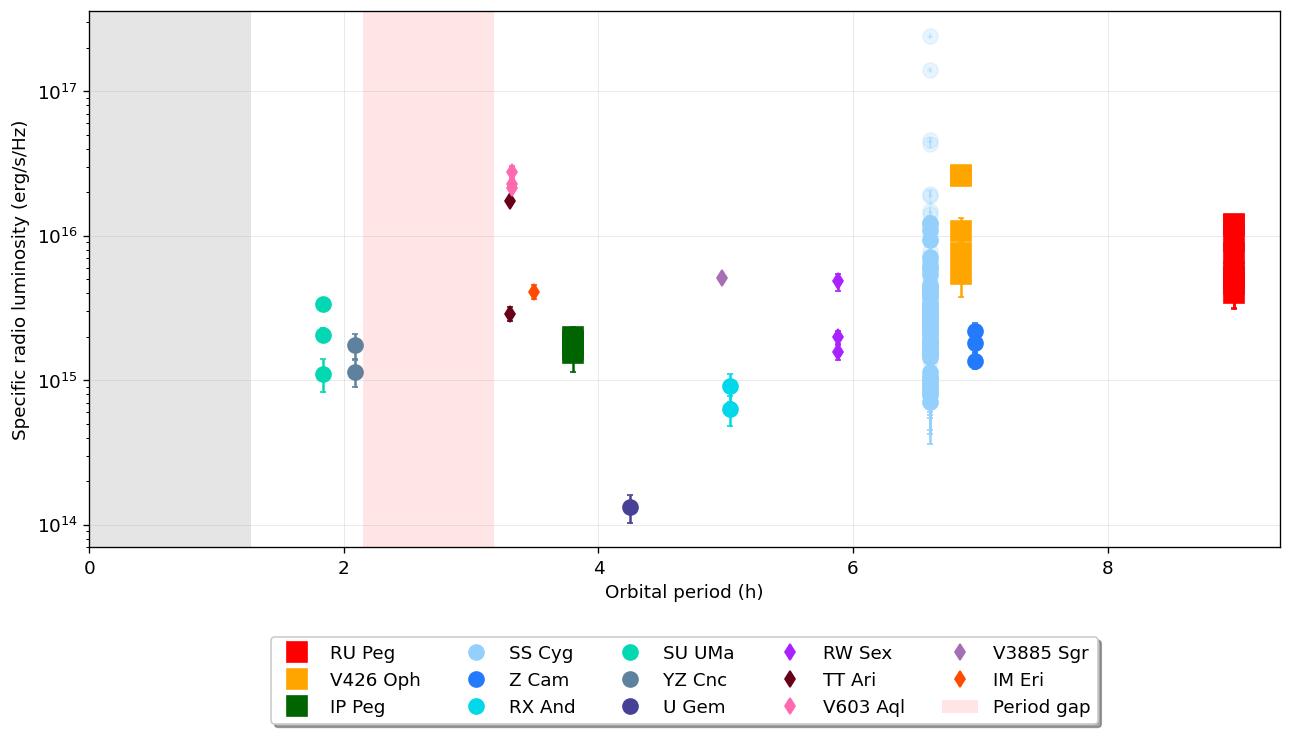}}
  \caption{Radio luminosity versus orbital period for detected DNe and novalikes. Novalikes are plotted with a smaller marker. For SS Cyg flare detections are plotted with a degree of transparency. In red is the period gap in which very few CVs are found. In grey is the area below the period minimum of 76.2 minutes \citep{kniggeDonorStarsCataclysmic2006}.}
  \label{fig:RadioVersusOrbitalPeriod}
\end{figure*}

To see if orbital period is an indicator for an object's radio luminosity, we show in Figure \ref{fig:RadioVersusOrbitalPeriod} radio luminosity as function of orbital period for the systems this article reports on, as well as for previously detected dwarf novae and novalikes. All dwarf novae detected after the 1980's are included. The brightest points of SS Cyg data correspond to radio flares, see e.g. \citet{mooleyRapidRadioFlaring2017,fenderLateoutburstRadioFlaring2019}. For DNe, higher orbital period roughly correlates with higher average accretion rate (mass transfer rate), as mentioned in the introduction.

The detected systems cover a wide range of orbital periods, both below and above the CV period gap. The divide lies between 2.15 and 3.18 hours \citep{kniggeDonorStarsCataclysmic2006}.

There is no discernible correlation of luminosity and orbital period and no significant luminosity difference between non-flaring DNe\footnote{The \emph{non-flaring} moniker is used to indicate that the top luminosity observations of SS Cyg are not considered for this statement. We classified these as flares. Which observations exactly are classified as such depends on the classification scheme used. We faded a number of SS Cyg data points in Figure \ref{fig:RadioVersusOrbitalPeriod} based on one such scheme. The top luminosity observations are clearly flares, and would be faded in any reasonable scheme.} in outburst and novalikes. But the sample of systems is small, and the long period DN systems RU Peg, V426 Oph and SS Cyg have a measured peak luminosity which is orders above that of the lower period systems SU UMa, YZ Cnc, IP Peg and U Gem. Concerning orbital period dependence of radio luminosity, we simply cannot make a determination either way. More data are needed.

IP Peg's luminosity is lower than that of RU Peg, V426 Oph and SS Cyg. This could be due to inclination. We observe IP Peg nearly edge on, as can be seen in Tables \ref{tab:PrevDNe} and \ref{tab:properties}. This could obscure the radio emission, most likely since it has to travel through the disc. Emission from the inner jet or inner disc are consistent with this, while emission from the secondary or from the intermediate region are not.

We do note considerable spread in radio luminosity, for the whole sample as well as for individual objects. For the DNe this is mostly related to their outburst phase. Novalikes may a priori be expected to have less variation in time of their radio luminosity, since they do not have outbursts and the disc is in the high state permanently. Still, optically it is known that also novalikes show a large spread in luminosity, on longer and more irregular time scales than DNe \citep[e.g.][]{Groot2001}. For some novalikes the luminosity spread in the radio band is as big as observed for some dwarf novae, so intrinsic variations are large.

With detections below the period gap, and a detection at the high end of the luminosity just above the period gap, for V603 Aql, it is safe to conclude that the current data give no reason to believe that radio emission luminosity varies systematically with orbital period. In a follow-up paper the significance of the non-detections will be discussed. No variation with orbital period translates to the statement that the average mass transfer rate (that is, the rate at which mass from the secondary enters the accretion disc) does not correlate with the radio luminosity. Furthermore, the radio luminosity is similar for DNe and NLs.

For V426 Oph the peak detection was at a luminosity of \qty{2.6e16}{erg.s^{-1}.Hz^{-1}}. This luminosity is the average over the one hour long observation. SS Cyg is known to sometimes have strong variations in radio flux during outburst, on timescales from days to minutes. A 15 minute flare was observed by \citet{mooleyRapidRadioFlaring2017}. By splitting the V426 Oph observation in 4 parts we can see if there is a short duration flare, or if all parts are together responsible for this average. We found that all parts are equally responsible.

Similarly, no variability on the scale of $\sim$15 minutes has been detected for IP Peg or RU Peg. Of the 6 dwarf novae studied in the previous two decades 4 have shown variability on such a time scale: SS Cyg, SU UMa, U Gem and RX And. Flaring during the rise to outburst was predicted in \citet{kordingTransientRadioJet2008}, based on the assumption that dwarf novae, like XRBs, produce a transient jet and that these objects behave similarly but scaled. SS Cyg not only showed flaring during the rise, but also a large 15 minute flare at the end of its radio outburst \citep{mooleyRapidRadioFlaring2017}. Such a flare was not observed for the three new sources, but it is entirely possible that a 15 minute large flare was simply missed in all cases, since we did only one observation per day, which means that for about 23 hours per day we did not observe.

\subsection{Radio luminosity versus optical luminosity}
 \begin{figure*}
  \resizebox{\hsize}{!}{\includegraphics{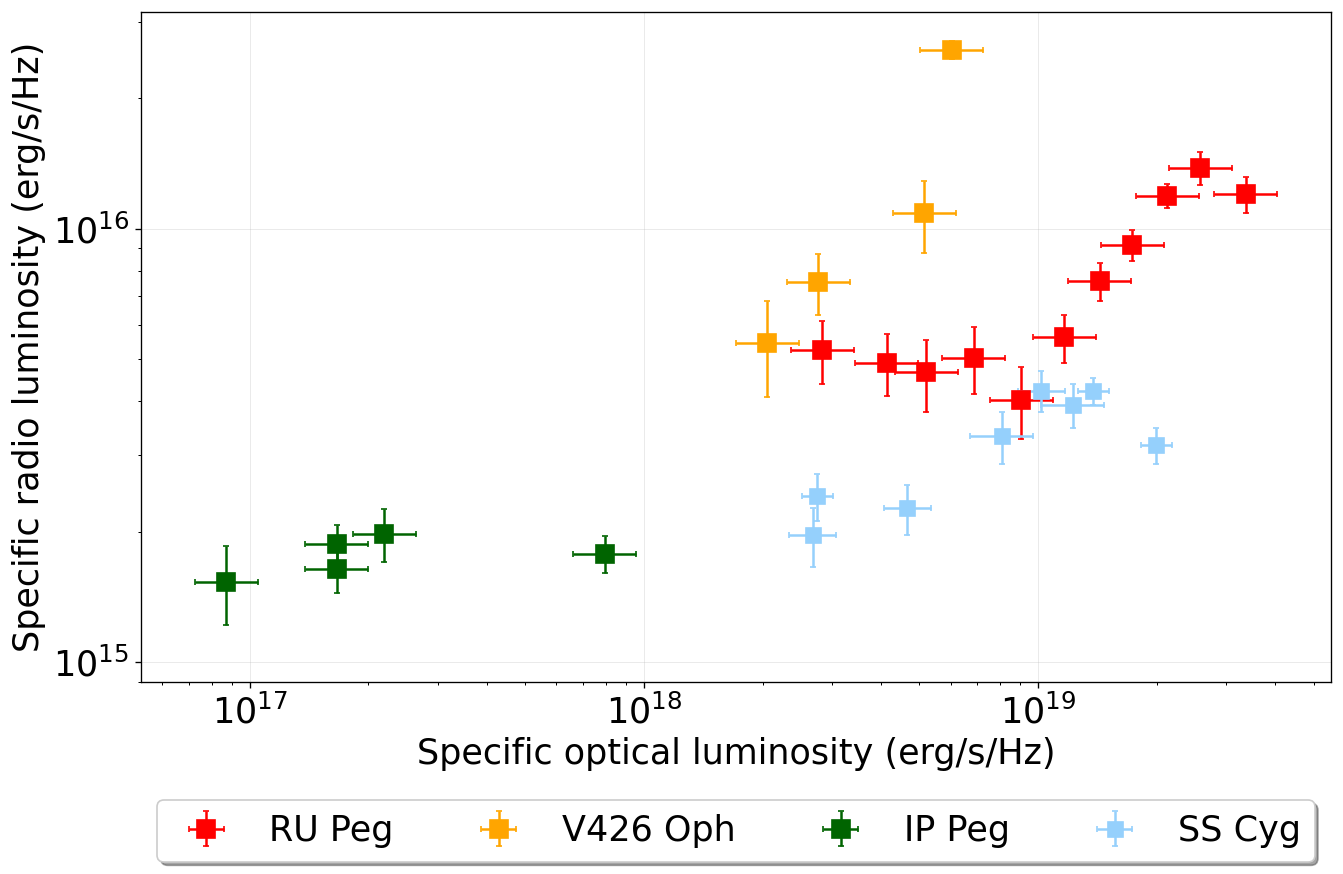}
  \hspace{1cm}
  \includegraphics{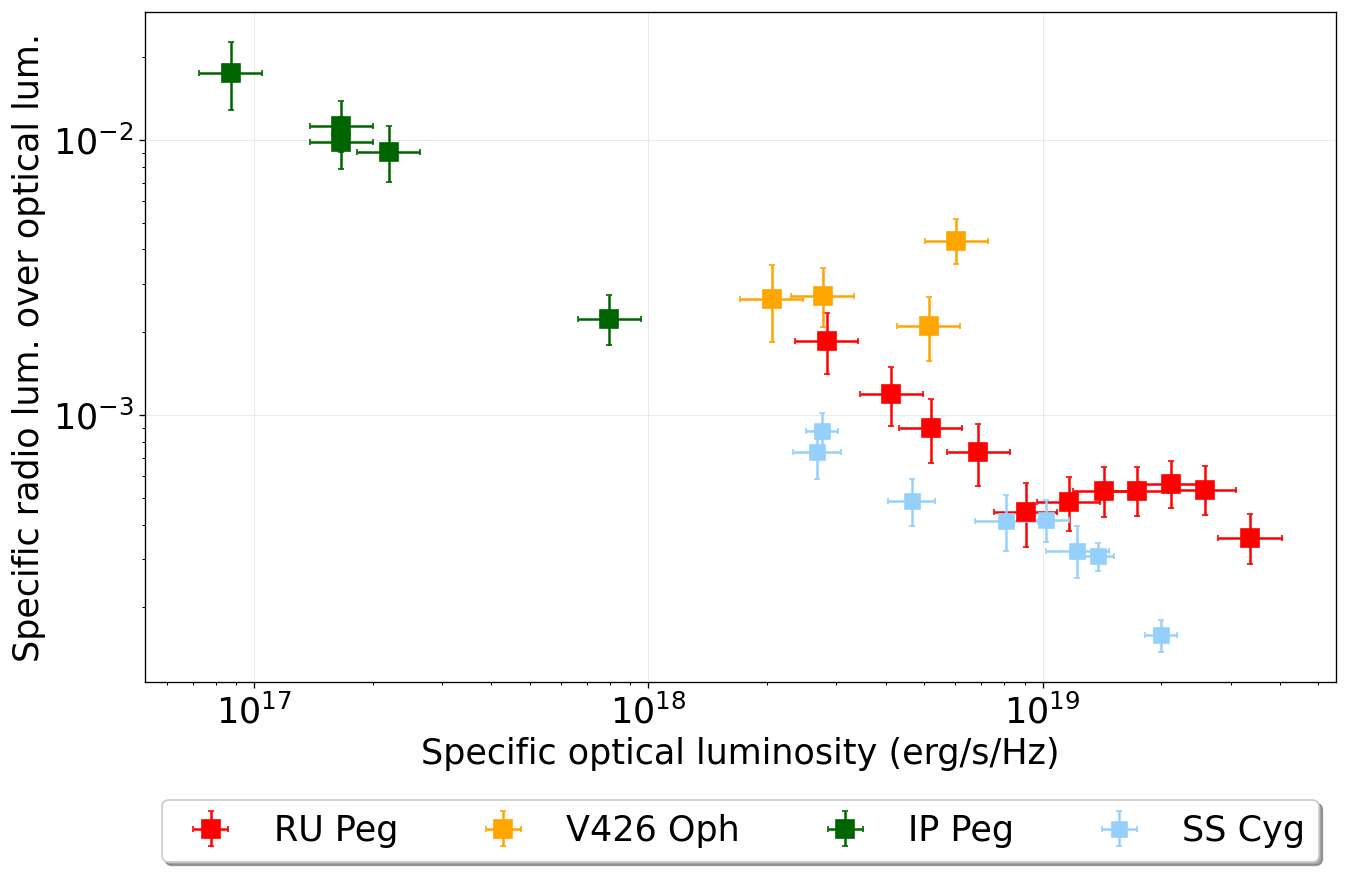}}
  \caption{{\it Left}Radio luminosity versus optical luminosity, not including upper limit measurements. Loglog plot. SS Cyg data from \citet{kordingTransientRadioJet2008} (8.6 GHz VLA) and \citet{miller-jonesInvestigatingAccretionDisk2010} (4.6 GHz VLA) are plotted for comparison. {\it Right} Ratio of specific radio luminosity over optical luminosity for the same objects.}
  \label{fig:RadioVersusOpticalLoglog}
\end{figure*}

Radio and optical emission come from different mechanisms and likely also from different regions. Outbursts of dwarf novae result in a time evolution of both types of radiation, which makes it possible to investigate a relation between radio luminosity and optical luminosity.

In Figure \ref{fig:RadioVersusOpticalLoglog} (left) we plot radio luminosity versus optical luminosity. The determined optical magnitude (based on \emph{V}, \emph{g} or similar bands, including visual determination by AAVSO astronomers) is assumed to be equal to the monochromatic AB magnitude. From the previously detected sources we plot only SS Cyg, since this is the only source with sufficient sampling. Combined, the three objects show a correlation, roughly a power law, but there is significant scatter around the relation, too much to make a strong statement. V426\,Oph is the most radio bright of the sample.

Individually, the objects seem to roughly follow a power law (straight line) relation between radio and optical flux.
Figure \ref{fig:RadioVersusOpticalLoglog} (right) shows the ratio of the radio over the optical luminosity.
A linear relation between radio and optical (fixed ratio) shows up as a horizontal line here. The figure shows that optically faint sources are relatively radio bright, and optical bright sources are relatively radio weak, reflecting the range of luminosities shown in the left hand panel of Fig. \ref{fig:RadioVersusOpticalLoglog}. The figure also shows that RU Peg is not exceptionally radio-bright even though it shows the highest optical luminosity.

The radio measurements of RU Peg during the optical decline show a linear radio-optical relation, forming a roughly horizontal line in Fig. \ref{fig:RadioVersusOpticalLoglog} (right). For IP Peg, V426 Oph and SS Cyg individually such roughly linear relation, horizontal in Fig. \ref{fig:RadioVersusOpticalLoglog} (right), could be present, at least for a large subset of points. However, taken as a sample together, a linear relation is clearly not present.

Radio emission can be expected to come from a small region near the white dwarf: the approximately 15 minutes that the very strong flare observed for SS Cyg lasted, indicates that it came from a small region. This radio emission could arise from a jet/outflow or from gyrosynchrotron emission near the white dwarf. Under this assumption, it is likely that the luminosity of the radio emission depends on the current accretion rate onto the white dwarf. UV and X-ray emission comes predominantly from the inner regions of the accretion disc and the boundary layer. In contrast, optical emission comes mainly from further out in the disc, and from the bright spot. This makes any power law relation between optical and radio luminosity measured at the same time hard to interpret. A delay (the travel time of matter from the main optical emission spot through the disc to its inner edge) should be taken into account and thus also the exact shape of the outburst. In \citet{schreiberDelaysDwarfNovae2003} the delay between optical outburst, UV outburst, EUV outburst and X-ray outburst for both inside-out (slow) rise and outside-in (double heating wave fast rise) is discussed. The time it takes for in-falling matter to make it from the outer disc to the white dwarf region depends on the effective viscosity ($\alpha$), which itself depends on the evolution of the outburst. In future work, it may be possible to model using the full shape of the outburst's optical lightcurve. From the theory side the DIM can provide the white dwarf accretion rate at any time. This is achieved by assuming that at each point in the disc and at all times the effective viscosity holds one of two values: the quiescence value and the outburst value. These two parameters are chosen as input for the DIM. See \citet{hameuryReviewDiscInstability2020}. The difficulty lies in having a DIM solution which describes the observed outburst exactly.

Concluding, we see a clear indication of a relation between optical and radio for dwarf nova outburst declines for the first time.

For the radio emission reported on in this article the origin and physical mechanism remain unknown. Although synchrotron radio emission from a jet near the white dwarf is the main candidate, we cannot fully rule out other possibilities, such as the candidate that was given as most likely for magnetic CVs in \citet{barrettRadioObservationsMagnetic2020}, namely that emission comes from the donor star (specifically EMCE from the lower corona). RU Peg's magnetically active donor and large irradiation are in favour of donor star emission. However, the fact that the radio emission is predominantly detected in outburst and that there is a clear radio decline with optical in RU Peg, shows that the radio emission mechanism is dependent on the accretion rate. This is strong evidence that not all radio emission is from a flaring secondary star.

RU Peg and IP Peg seem to show emission even during quiescence (see Table \ref{tab:luminosities}). For V426 Oph this is not detected, but a nearby bright source makes such detection harder for this system. It is possible that the mechanism responsible for the emission during the outburst does not fully go quiet during quiescence. It is also possible that there are multiple mechanisms at play, but this would mean that both mechanisms cause radio emission of similar order of luminosity. One of these mechanisms is then closely following the optical outburst, and therefore closely following the enhanced accretion rate, the other is not varying or varying much slower. For the quiescence emission it cannot fully be excluded that the donor star alone is responsible. Single stars of the donor star spectral type have been observed to emit in radio. \citet{gudelRadioXrayEmission1992} suggests that there is a population of strongly emitting K dwarfs and M dwarfs, and a population of much less strongly emitting dwarfs. The radio emission correlates with X-ray emission, which indicates coronal activity, and with rotational period, with faster rotation resulting in more radio emission,although this effect saturates for fast rotation. The proposed emission mechanism is gyrosynchrotron emission of electrons trapped in magnetic loops in the corona. This single star quiescent emission can reach levels close to what is observed here, with \citet{gudelRadioXrayEmission1992} for example measuring $L_\text{R} > 10^{15} $\unit{erg.s^{-1}.Hz^{-1}} for a K5V star. However, there are many nearby K dwarfs (over 5000 within 50 parsec) and M dwarfs, with only a very small fraction of them being detected in radio.

\subsection{Conclusion}
We detected three new dwarf novae in radio. With this we increased the number of detected dwarf novae at and below L-band frequency from one to four. These detections also added three objects to the general list of radio-detected dwarf novae, bringing the total to ten (conservatively excluding the likely source-confused ones from \citet{turner12cmObservationsStellar1985} and the EM Cyg detection from \citet{benzVLADetectionRadio1989}, done in the 1980's). These three new dwarf novae were all detected again during a later outburst, showing that radio detection was consistent for these sources. The maximum radio luminosity for the dwarf nova population and for the novalike population was found to be similar. Detections of both source types are found for orbital periods all over the range, which goes against the a priori expectation that longer orbital period corresponds to more emission. For V426 Oph we found a very high radio luminosity: the highest radio luminosity ever for a dwarf nova outside a flaring period. The radio peak for the newly detected dwarf novae was found around the time of the optical peak. A smooth decline during the optical decline phase of the outburst was observed for V426 Oph and, most prominently, for RU Peg.

\section*{Acknowledgements}
PJG is partly supported by NRF SARChI Grant 111692. PAW kindly acknowledges support from the University of Cape Town and the NRF.

The MeerKAT telescope is operated by the South African Radio Astronomy Observatory, which is a facility of the National Research Foundation, an agency of the Department of Science and Innovation. We thank everyone who has been involved in the design, building and operation of this telescope. We kindly thank Sarah Buchner for the assistance with the dynamic scheduling of the MeerKAT observations during the respective outbursts.

We acknowledge the use of the Ilifu cloud computing facility – www.ilifu.ac.za, a partnership between the University of Cape Town, the University of the Western Cape, Stellenbosch University, Sol Plaatje University and the Cape Peninsula University of Technology. The Ilifu facility is supported by contributions from the Inter-University Institute for Data Intensive Astronomy (IDIA – a partnership between the University of Cape Town, the University of Pretoria and the University of the Western Cape, the Computational Biology division at UCT and the Data Intensive Research Initiative of South Africa).

This work has made use of data from the European Space Agency (ESA) mission {\it Gaia} (\url{https://www.cosmos.esa.int/gaia}), processed by the {\it Gaia} Data Processing and Analysis Consortium (DPAC, \url{https://www.cosmos.esa.int/web/gaia/dpac/consortium}). Funding for the DPAC has been provided by national institutions, in particular the institutions participating in the {\it Gaia} Multilateral Agreement.

%%%%%%%%%%%%%%%%%%%%%%%%%%%%%%%%%%%%%%%%%%%%%%%%%%
\section*{Data Availability}

The uncalibrated MeerKAT visibility data presented in this paper are available in the archive of the South African Radio Astronomy Observatory at https://archive.sarao.ac.za. The continuum MeerKAT observations were taken as part of the ThunderKAT Large Survey programme, project code SCI-20180516-PW-02. Data that are not available through public archives, and all source codes, will be shared on reasonable request to the corresponding author.


\bibliographystyle{mnras}
\bibliography{Bibliography} % The bibtex file is called Bibliography.bib

%%%%%%%%%%%%%%%%%%%%%%%%%%%%%%%%%%%%%%%%%%%%%%%%%%

%%%%%%%%%%%%%%%%% APPENDICES %%%%%%%%%%%%%%%%%%%%%

%%%%%%%%%%%%%%%%%%%%%%%%%%%%%%%%%%%%%%%%%%%%%%%%%%

% Don't change these lines
\bsp	% typesetting comment
\label{lastpage}
\end{document}